\documentclass[preprint,showpacs]{revtex4}
\usepackage{graphicx}
\usepackage{bm}
\usepackage{dcolumn}
\usepackage{amsmath}
\usepackage{lscape}
\usepackage{longtable}
\usepackage{rotating}

\def \lleq {\lower0.9ex\hbox{ $\buildrel < \over \sim$} ~}
\def \ggeq {\lower0.9ex\hbox{ $\buildrel > \over \sim$} ~}

\def \beq  {\begin{equation}}
\def \eeq  {\end{equation}}
\def \ber  {\begin{eqnarray}}
\def \eer  {\end{eqnarray}}

\begin{document}
\newcommand{\newc}{\newcommand}

\newc{\be}{\begin{equation}}
\newc{\ee}{\end{equation}}
\newc{\ba}{\begin{eqnarray}}
\newc{\ea}{\end{eqnarray}}
\newc{\bea}{\begin{eqnarray*}}
\newc{\eea}{\end{eqnarray*}}
\newc{\D}{\partial}
\newc{\ie}{{\it i.e.} }
\newc{\eg}{{\it e.g.} }
\newc{\etc}{{\it etc.} }
\newc{\etal}{{\it et al.}}
\newcommand{\nn}{\nonumber}
\newc{\ra}{\rightarrow}
\newc{\lra}{\leftrightarrow}
\newc{\lsim}{\buildrel{<}\over{\sim}}
\newc{\gsim}{\buildrel{>}\over{\sim}}
\title{On the Dark Sector Interactions}
\author{Rong-Gen Cai}
\email{cairg@itp.ac.cn}
\author{Qiping Su}
\email{sqp@itp.ac.cn}
\address{
Key Laboratory of Frontiers in Theoretical Physics, Institute of
Theoretical Physics, Chinese Academy of Sciences, P.O. Box 2735,
Beijing 100190, China}

\date{\today}

\begin{abstract}
It is possible that there exist some interactions between dark
energy (DE) and dark matter (DM), and a suitable interaction can
alleviate the coincidence problem. Several phenomenological
interacting forms are proposed and are fitted with observations in
the literature. In this paper we investigate the possible
interaction in a way independent of specific interacting forms by
use of observational data (SNe, BAO, CMB and Hubble parameter). We
divide the whole range of redshift into  a few bins and set the
interacting term $\delta(z)$ to be a constant in each
redshift bin. We consider four parameterizations of the equation of
state $w_{de}$ for DE and find that $\delta(z)$ is likely to cross
the non-interacting ($\delta=0$) and have an oscillation form. It suggests that to study the
interaction between DE and DM, more general phenomenological forms
of the interacting term should be considered.
\end{abstract}


\maketitle

\section{Introduction}

It has been suggested from astronomical observations that the main
components of our universe are dark matter (DM) and dark energy
(DE). DM  behaves like the usual baryon matter and can form
clusters, while DE is uniformly distributed in the whole universe,
and it derives the universe to accelerating expand. Very ironically,
we have known a little on DM and DE so far. The questions of what
particles DM's are and what nature DE is remain open. From
astronomical observations, however, some properties of DE can be
deduced. For example, usually one characterizes DE with its equation
of state $w_{de}$, the ratio of the pressure to the energy density
of DE; $w_{de}$ is found very close to $-1$ from the observations.
Therefore a natural candidate of DE is the well-known cosmological
constant introduced by Einstein in 1917, for which the equation of
state is exact $-1$. Although the cosmological constant is a
beautiful and economic candidate, it suffers from some theoretical
puzzles to be explained as currently observed DE. The theoretical
difficulties (puzzles) are so-called fine-tuning problem and
coincident problem (i.e., why energy densities of DE and DM happen
to be of the same order today?)

To avoid these problems, some dynamical DE models have also been
proposed in the literature. The simplest dynamical DE model is a
time-dependent scalar field. Based on different forms of the
Lagrangian of scalar field, the scalar field models could be
classified into quintessence, K-essence, phantom and quintom models.
Furthermore, due to the ignorance for DM and DE, one is not sure
whether there exists any direct interaction between DM and DE, at
least no known symmetries prevent such interaction. Indeed, possible
interactions between DM and DE have been intensively investigated in
recent years.  It has been shown that a suitable interaction can
help to alleviate the coincidence problem
\cite{Amendola:1999er,Chimento:2003iea,Cai:2004dk,Olivares:2006jr}.
Various interacting models have been studied
\cite{Chimento:2009hj,He:2008tn,Szydlowski:2005kv,Bean:2008ac,Chimento:2007yt,Feng:2007wn,Micheletti:2009pk,Chen:2008ca,Wei:2007ut,Wei:2007zs,Quartin:2008px}.
Several phenomenological interacting forms have been proposed and
have been fitted with observations
\cite{Jesus:2008xi,Valiviita:2009nu,Zimdahl:2002zb,Boehmer:2008av,Zimdahl:2001ar,Guo:2007zk,Mangano:2002gg}.
Some recent discussions seemingly imply that the decaying of DM into
DE is favored \cite{Costa:2009mv,Pereira:2008at} by observations,
which can make the coincidence problem more severe. However, most of
those studies depend on the interacting forms, that is to say, those
results are obtained by taking some special interacting terms. In
other words, those studies are model dependent. Moreover, most of
the models exclude the possibility of an oscillation interaction.
And if the interaction exists, by fitting, one could only conclude
that either DM decays to DE, or DE decays to DM.

In this paper, we  are going to study the interaction in a way
independent of the interacting form by observational data. To do that,
we divide the whole redshift range into a few bins and the
interacting term $\delta(z)$ is set to be a constant in each bin
\cite{Huterer:2004ch,Sullivan:2007pd}. Clearly such study depends on
 DE models and the number of bins.
We will study 3-6 bins cases with a constant $w_{de}$ and try to get some common features of $\delta(z)$.
To see effect for different DE models,  we will adopt four
different parameterizations of $w_{de}$ with a preferable division of bins.
We will fit the interacting models with the Union SnIa \cite{Kowalski:2008ez}, BAO
\cite{Percival:2009xn}, 9 Hubble data \cite{Simon:2004tf} and the
shift parameter R from WMAP5 \cite{Komatsu:2008hk}.  We obtain the
best-fitted parameters and likelihoods by using the MCMC method. We
find that $\delta(z)$ is likely to be oscillating and to cross the
non-interacting ($\delta=0$) line. We  also compare behaviors of
$r=\rho_m/\rho_{de}$ in the best-fitted models with those of
corresponding models without interaction.  In  three cases of four
parameterizations of $w_{de}$, the coincidence problem is
alleviated, though DM decays into DE in some regions of redshift.

\section{Methodology}
We  consider interacting models in a flat FRW universe
 \be
3H^2=\rho_\gamma+\rho_b+\rho_{de}+\rho_{dm}, \ee where $\rho_\gamma$
and $\rho_b$ are energy densities of radiation and baryon,
respectively, and $\rho_{de}$ and $\rho_{dm}$ are energy densities
of DE and DM, respectively.  We have set the speed of light $c=1$
and $8\pi G=1$. The continuity equations for energy densities of the
interacting DM and DE are \ba
\dot{\rho}_{dm}+3H\rho_{dm}=3H\delta, \nonumber\\
\dot{\rho}_{de}+3H(1+w_{de})\rho_{de}=-3H\delta. \label{eoi} \ea In
some phenomenological models of interaction, the interacting term
$\delta$ is always assumed to be a function of $\rho_{dm}$ and $
\rho_{de}$, such as $\delta=\lambda\rho_{dm}$
\cite{Zimdahl:2001ar,Guo:2007zk}, $\delta=\lambda\rho_{de}$
\cite{Pavon:2005yx,Pereira:2008at} or
$\delta=\lambda(\rho_{dm}+\rho_{de})$ \cite{Chimento:2003iea}, thus
the constraints resulting from observations will depend on the form
of $\delta$. By fitting, if $\lambda=0$, it indicates that there
does not exist interaction between DM and DE; if $\lambda >0$, it
stands for the decay direction from DE to DM; while from DM to DE,
if $\lambda <0$.  However, obviously the way loses the possibility
that $\delta$ has an oscillating behavior.

To investigate such a possibility, in this paper we divide the whole
redshift into four bins and  set $\delta$ to be a piecewise constant
in each redshift bin
 \be \delta(z_{n-1}<z\leq z_n)=\delta_n, \
(n\geq1) \label{bd} \ee In our main analysis, we will set $z_0=0$,
$z_1=0.2$, $z_2=0.5$, $z_3=1.8$ and $z_4=1090$. Also we will
consider possible effect of the number of bins on the fitting
results.

From Eq. (\ref{eoi}) we have \ba
&& \rho_{dm}(z)=\rho_{dm}^0(1+z)^3-(1+z)^3\int^z_0\frac{3\delta(x)}{(1+x)^4}dx, \nonumber\\
&&
\rho_{de}(z)=\rho_{de}^0F(z)+F(z)\int^z_0\frac{3\delta(x)}{(1+x)F(x)}dx,\label{et}
\ea where $F(z)=\exp[\int^z_0\frac{3(1+w_{de})}{1+x}dx]$ and
superscript 0 represents the present value. As
$\rho_b=\rho_b^0(1+z)^3$, we can write $\rho_{dm}$ and $\rho_b$
together as $\rho_m$. With the piecewise constant $\delta(z)$ we
have an analytical form for $\rho_m$
 \be \rho_{m}(z_{n-1}<z\leq
z_n)=[\rho_{m}^0-\delta_1+\sum_{i=1}^{n-1}(\delta_i-\delta_{i+1})(1+z_i)^3](1+z)^3+\delta_n
\ee
 where $\rho_m^0=\rho_b^0+\rho_{dm}^0$.

For energy density $\rho_{de}$ of DE, we will employ four different
parameterizations of $w_{de}$ as follows.

I. $ \ w_{de}=-1 $.  In that case, $\rho_{de}$ can be written
analytically as
 \be \rho_{de}(z_{n-1}<z\leq
z_n)=\rho_{de}^0+3\sum_{i=1}^{n-1}(\delta_i-\delta_{i+1})\ln(1+z_i)+3\delta_n\ln(1+z)\label{al}
\ee

II. $ \ w_{de}=w_0 $. In this case, we have
  \ba && \rho_{de}(z_{n-1}<z\leq z_n)= [\rho_{de}^0+\frac{\delta_1}{1+w_0}\nonumber \\
&&~~~~~
-{1\over(1+w_0)}\sum^{n-1}_{i=1}(\delta_i-\delta_{i+1})(1+z_i)^{-3(1+w_0)}](1+z)^{3(1+w_0)}
-\frac{\delta_n}{(1+w_0)}\label{ac}
\ea

III. $ w_{de}=w_0+w_1z/(1+z)$. In this case, $F(z)$ in Eq.
(\ref{et}) has the form \cite{Chevallier:2000qy,Linder:2002et}
 \be
F(z)=(1+z)^{3(1+w_0+w_1)}\exp(-\frac{3w_1z}{1+z}) \ee

IV. $w_{de}=w_0+w_1z/(1+z)^2 $. In this case $F(z)$ can be expressed
as
 \be
F(z)=(1+z)^{3(1+w_0)}\exp(\frac{3w_1z^2}{2(1+z)^2}) \ee

For parameterizations III and IV, it is hard to get analytic forms
of $\rho_{de}$ as Eqs. (\ref{al}) and (\ref{ac}). Now the Friedmann
equation of the interacting models can be written as: \be
E^2(z)=\frac{H^2(z)}{H_0^2}=\Omega_r^0(1+z)^4+\rho_{m}/3H_0^2+\rho_{de}/3H_0^2
\ee

We now fit these four models with observations. The observational
data to be used are the 307 Union SNIa data \cite{Kowalski:2008ez},
the Baryon Acoustic Oscillation (BAO) data from SDSS DR7
\cite{Percival:2009xn}, the shift parameter R from WMAP5
\cite{Komatsu:2008hk}, and 9 data of the Hubble parameter $H(z)$
\cite{Simon:2004tf}. We obtain the best-fitted parameters by
minimizing \be
\chi^2_{tot}=\widetilde{\chi}^2_{sn}+\chi^2_{bao}+\chi^2_{R}+\chi^2_{H}+(h-0.742)^2/0.036^2
\ee where  $h=H_0/100 {\rm km \cdot s^{-1}\cdot Mpc}^{-1}$. We
acquire the constraints by using the MCMC method.

For 307 Union SNIa data, $\chi^2_{sn}$ is defined as
 \be
\chi^2_{sn}=\sum_i\frac{[\mu_{th}(z_i)-\mu_{ob}(z_i)]^2}{\sigma^2(z_i)}\label{chi}
\ee where $\mu_{th}(z)=5\log_{10}[(1+z)\int_0^zdx/E(x)]+\mu_0$, and
$\mu_0=42.384-5\log_{10}h$ is a nuisance parameter. One can expand
Eq. (\ref{chi}) as
$$\chi^2_{sn}=A+2\mu_0B+\mu_0^2C$$
where
\ba
A &=&\sum_i\frac{[\mu_{th}(z_i;\mu_0=0)-\mu_{ob}(z_i)]^2}{\sigma^2(z_i)}\  ,\nonumber\\
B &=&
\sum_i\frac{\mu_{th}(z_i;\mu_0=0)-\mu_{ob}(z_i)}{\sigma^2(z_i)} , \
C=\sum_i\frac{1}{\sigma^2(z_i)} \ea
 We adopt the minimization of
$\chi^2_{sn}$ with respect to $\mu_0$ to replace $\chi^2_{sn}$
$$\widetilde{\chi}^2_{sn}=\chi^2_{sn,min}=A-B^2/C$$
In fact, it is equivalent to performing an uniform marginalization over $\mu_0$ \cite{Nesseris:2005ur}.

For the BAO data, one has
\be
\chi^2_{bao}=\frac{[D_V(0.35)/D_V(0.2)-1.736]^2}{0.065^2} \ee where
$D_V(z)=[z/H(z)(\int_0^zdx/H(x))^2]^{1/3}$.

For the shift parameter, we take
 \be \chi^2_R=\frac{(R-1.71)^2}{0.019^2}
\ee where $R=\sqrt{\Omega_m^0}\int_0^{z_*}\frac{dz}{E(z)}$ and
$z_*=1090$.

And for the Hubble evolution data, we have
 \be
\chi^2_H=\sum_{i=1}^9\frac{[H(z_i)-H_{ob}(z_i)]^2}{\sigma_i^2} \ee
Note that we also have used a Gaussian prior $h=0.742\pm0.036$
\cite{Riess:2009pu}.

\section{Results}

\begin{table}[b]
\caption{The best-fitted values of $\delta_i$ in $z\in(0,1.8)$ for 3-6 bins.
The values in () is the upper boundary of the redshift bin.}\label{bf}
\begin{tabular}{c|cccccc}
 \hline
 \hline
Models~&~$\delta_1$~&~$\delta_2$~&~$\delta_3$~&~$\delta_4$~&~$\delta_5$~&~$\delta_6$\\
 \hline
 3 bins&-0.32~(0.6)&0.12~(1.2)&5.80~(1.8)&-&-&-\\
 4 bins&-0.17~(0.45)&-0.69~(0.9)&5.13~(1.35)&-2.54~(1.8)&-&-\\
 5 bins&0.05~(0.36)&-1.09~(0.72)&0.16~(1.08)&10.12~(1.44)&-10.32~(1.8)&-\\
 6 bins&0.09~(0.3)&-1.84~(0.6)&0.29~(0.9)&6.11~(1.2)&1.08~(1.5)&62.20~(1.8)\\
 \hline
 \hline
\end{tabular}
\end{table}

Now we fit our models with the observations. As the data only give
very weak constraint for $z>1.8$, we fix
$\delta(1.8<z<1090)=0$ in our main analysis. To obtain the
constraint for a specified parameter, we marginalize over all other
parameters by using the MCMC method. In addition, in all
computations, we demand that $\rho_{de}$ and $\rho_m$ keep positive
in the range of $z\in(0, 1090)$. We do not decorrelate the
constraints in the different redshift bins. The constraints of
$\delta_i$ are correlated. But in this way it ensures that the
constraints obtained for a given bin are confined to the exact
redshift range of the bin, as discussed in
\cite{dePutter:2007kf,Kowalski:2008ez}. As the fitting results might
depend on the divided method of redshift bins and the dark energy
models, we will study two situations:

A. different numbers of bins with a constant $w_{de}$;

B. four different parameterization of $w_{de}$ with a preferable division of bins.

\subsection{Effects of the number of bins}

\begin{figure}[b]
  \includegraphics[width=7.0in,height=4.5in]{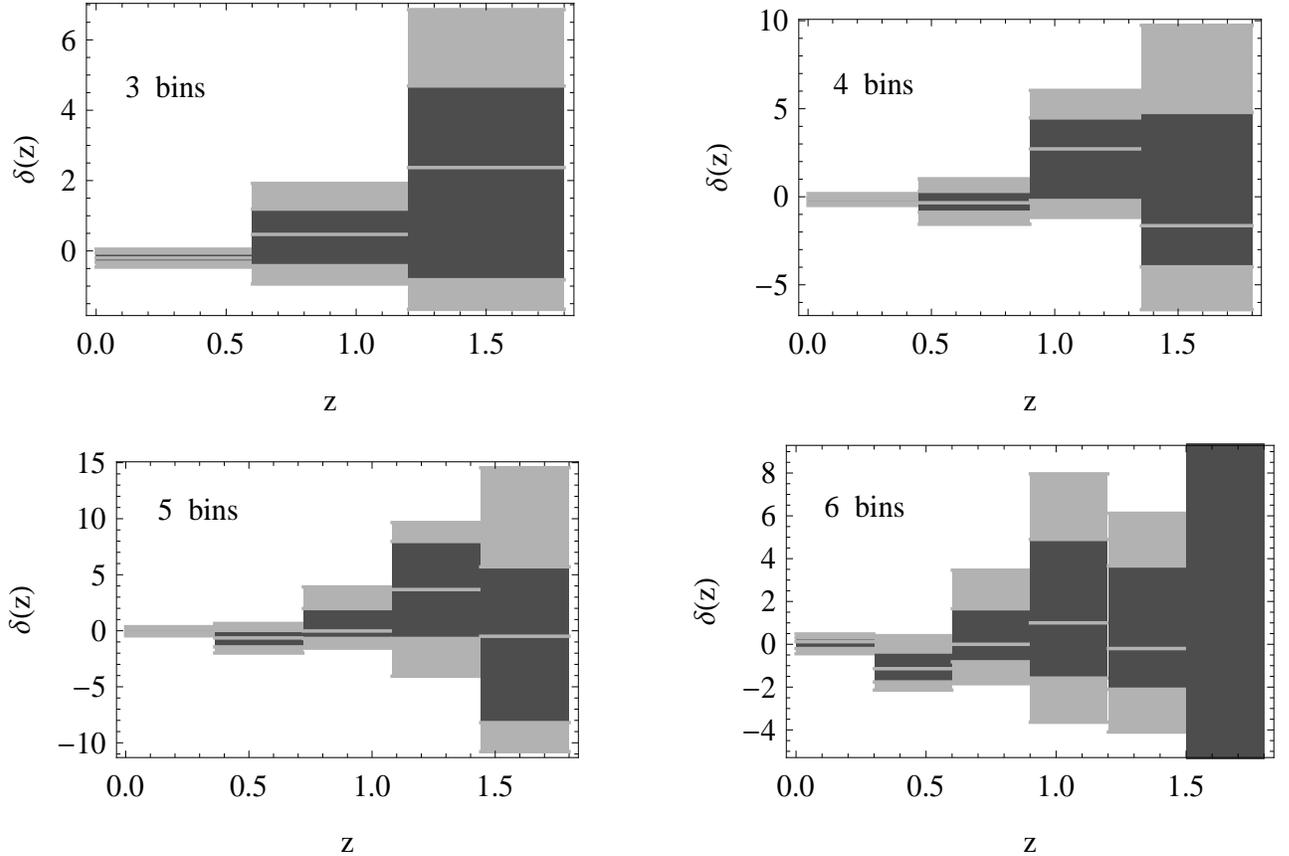}
  \caption{Constraints of $\delta(z)$ at 68\% and 95\% c.l. for 3, 4, 5 and 6 bins in $z\in(0,1.8)$.
  The bins are equally divided.
The equation of state of DE is assumed as a constant.
  $\delta(z)$ is in the unit of $3H_0^2$.}\label{dbins}
\end{figure}

At first, to see effects of the number and locations of the redshift
bins, we divide the region of $z\in(0,1.8)$ equally into 3, 4, 5 and
6 bins respectively and assume $w_{de}$ to be a constant. The
best-fitted $\delta_i$ from the observations are shown in
Table~\ref{bf}. In the most
cases the result prefers that $\delta(z)$ crosses the
non-interaction line ($\delta=0$) around $z=0.5$ and there is likely an oscillation of the interaction term
$\delta(z)$. The corresponding errors of $\delta(z)$ are shown in
the Fig.~\ref{dbins}. In 95\% c.l., all results for 3-6 bins show
that $\delta$ is consistent with $0$ from the observations. But in
68\% c.l., $\delta(z)$ is minus at the first bin ($z\lesssim0.5$) in
the 3 and 4 bins cases and at the second bin (around $z=0.5$) in the
5 and 6 bins cases . If one uses other three parameterizations of DE
that introduced in the section II, one would get similar results,
i.e., there is always a bin in which $\delta(z)$ departs from 0 in
68\% c.l. . As a result we may conclude from Fig.~\ref{dbins} that

1. With more bins, more finer structure of $\delta$ can be resolved,
e.g., for more than 3 bins oscillation behaviors of $\delta$ appear.
But for more bins the constraints of $\delta(z)$ in each bin from the observations will be weaker.

2. The errors for $z>1$ are much bigger than that for $z\in(0,1)$.
It is mainly due to the fact that there are much less data points in the large redshift region.

It is also very likely that $\delta(z)$ crosses the $\delta=0$ line around $z=0.5$.

\subsection{Effects of parameterizations of DE}

\begin{figure}[b]
  \includegraphics[width=7.0in,height=4.5in]{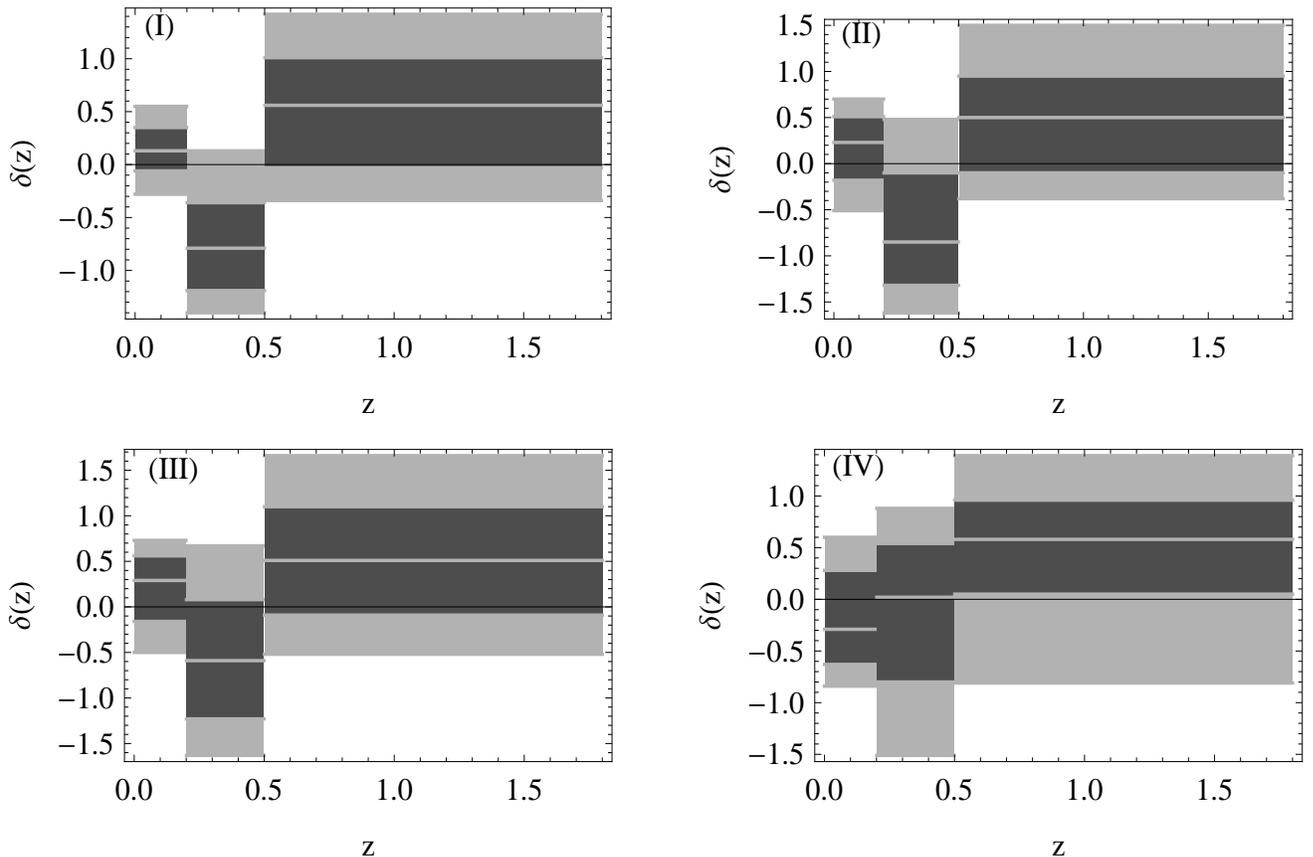}
  \caption{The constraints of $\delta_i$ at 68\% and 95\% c.l. . $\delta_i$ is in the unit of $3H_0^2$.} \label{delta}
\end{figure}

\begin{figure}[t]
  \includegraphics[width=7.0in,height=4.5in]{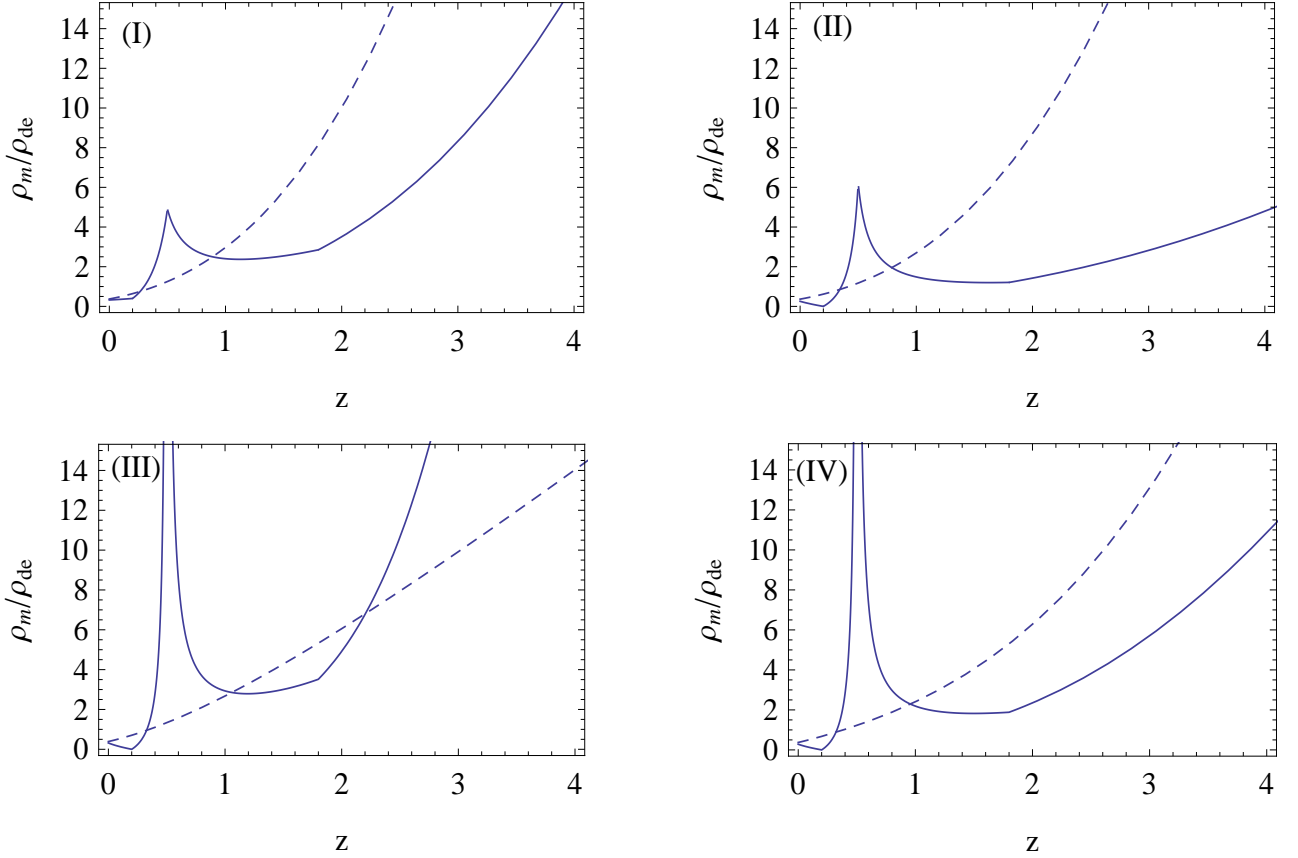}
  \caption{The behaviors of $r=\rho_m/\rho_{de}$.
 Solid curves are for the best-fitted models shown in Table \ref{ta} (with
 $\delta_4=0$), while the
 dashed curves  for the best-fitted models with $\delta(z)=0$.} \label{ratio}
\end{figure}

By considering these conclusions, in what follows we will divide the
region of $z\in(0, 1.8)$ into three bins as: $(z_0=0, z_1=0.2,
z_2=0.5, z_3=1.8)$~\cite{Sullivan:2007pd}, which is the case
adopted by most discussions in the literature and from which
fine constraints of $\delta(z)$ could be obtained indeed.
Four parameterizations of DE introduced in section II will be used.
The best-fitted parameters and the constraints at 68\% and 95\%
c.l. are shown in Table \ref{ta}. The best-fitted parameters for models
with $\delta_4$ unfixed are also shown in Table \ref{ta}. There are
almost no differences between the models with $\delta_4$ fixed and
unfixed.  The corresponding 68\% and 95\%
constraints are shown in Fig.~\ref{delta}, and the Fig.~\ref{ratio}
shows the behaviors of the ratio $r=\rho_m/\rho_{de}$ in the
best-fitted models with interaction, compared with the cases
without interaction.

I. $w_{de}=-1$

From Table \ref{ta} and  Fig.~\ref{delta}, it is obvious that $\delta_2$
is negative, while $\delta_1$ and $\delta_3$ are positive, which
implies that $\delta(z)$ could cross the $\delta=0$ line at a
recent time and the decay direction between DM and DE could be
variable. When $\delta(z)$ crosses the non-interacting line, the
effective equation of state of DE will cross the cosmological
constant ($w=-1$) line. Thus DE behaves as a quintom fluid. As
shown from Table \ref{ta} and Fig.~\ref{delta}, $\delta_2$ departs from
0 beyond 68\% c.l. But  models (III) and (IV), which have more
degrees of freedom, show that  it is still consistent with
$\delta=0$ everywhere in 68\% c.l. . In addition,  it can be seen
from Fig.~\ref{ratio} that the interaction can help to alleviate
the coincidence problem in this case.

II. $w_{de}=w_0$

For this parametrization, the situation is similar to the case of
$w_{de}=-1$. The sign of $\delta(z)$ can be varied in the different
bins. The possibility of $\delta_2<0$ is larger than 68\%, but less
than 95\%. The coincidence problem is also alleviated in the
best-fitted model.

 III. $ w_{de}=w_0+w_1z/(1+z)$

In this case, there is still a downward departure of $\delta_2$
from $0$, but now the constraint is consistent with $\delta(z)=0$
in 68\% c.l. It looks from Fig.~3 that the coincidence problem
could not be alleviated in this case. Note that to avoid a serious
degeneracy, we have assumed a prior $\Omega_m<0.37$ here.

 IV. $w_{de}=w_0+w_1z/(1+z)^2 $

In this case, $\delta(z)=0$ is consistent with the observations in
68\% c.l.  There is still a possibility of crossing the
non-interacting line. The coincidence problem can be alleviated.

As expected, the resulting constraints are effected by
parameterizations of $w_{de}$ and divisions of bins. But for all
cases we have considered here we see that the interacting term
prefers to have a behavior crossing the non-interacting line, and
there might exist an oscillation $\delta(z)$ in the most cases.

\section{Conclusion}
We have investigated the constraints of the interaction between DE
and DM from the observational data. To make the constraints
independent of  specific interacting forms, we divide the whole
redshift into four bins. In each bin $\delta(z)$ is set to be a
constant $\delta_i$.
First we have estimated effects of numbers of bins by consider 3-6 bins with a constant $w_{de}$,
from which we get some common features and choose a preferable division of bins:
$(z_0=0, z_1=0.2, z_2=0.5, z_3=1.8)$.
For models of DE, we have adopted four
parameterizations of $w_{de}$. The resulting constraints of
$\delta_i$ depend on these parameterizations. But there are also some common
features of the interaction. The results are
summarized as follows.

1. The observational data prefer that $\delta(z)$ crosses the
$\delta=0$ line and has an oscillation behavior at a recent time,
which implies that the decay direction can be variable. It is
similar to the case that the equation of state of DE is likely to
cross the cosmological constant ($w=-1$) line. For many well studied
phenomenological interacting forms, such as
$\delta=\lambda\rho_{dm}$ and $\delta=\lambda(\rho_{dm}+\rho_{de})$,
the sign of $\delta(z)$ is unchangeable.  Our results raise the
possibility that $\delta(z)$ can have different signs at the
different times. It implies that more general phenomenological forms
of the interaction should be considered, if the interaction indeed
exists.

2. The constraints given from observations show a departure of
$\delta(z)$ from 0 beyond 68\% c.l. for the $w_{de}=-1$ and
$w_{de}=w_0$ parameterizations. But for other two parameterizations
of DE, $w_{de}=w_0+w_1z/(1+z)$ and $w_{de}=w_0+w_1z/(1+z)^2$ which
have more degrees of freedom, the constraints are consistent with
$\delta(z)=0$. To confirm the existence of the interaction, more
observations and theoretical studies are needed.

3. The coincidence problem can be alleviated in the three cases of
four models, compared to corresponding ones without the interaction.
The ratio $r=\rho_m/\rho_{de}$ will evolve more rapidly (slowly)
when $\delta(z)<0$ ( $\delta(z)>0$ ) than the cases without the
interaction. The decay of DE to DM ($\delta(z)>0$) can alleviate the
coincidence problem, while the decay of DM to DE ($\delta(z)<0$)
will make it more severe. Though $\delta(z)$ is negative somewhere
in the best-fitted models, its effects can be offset by that of the
$\delta(z)>0$ regions and the period of $r\sim O(1)$ can be longer
than that of the corresponding non-interacting models. This way the
coincidence problem is alleviated.

Due to the ignorance on the properties of DE and DM, to study the
dark sector interactions one always needs to assume some models of
DE and DM. DM is always assumed as pressureless fluid, while there
exist plenty variants of DE models. In the sense of phenomenology,
the main difference among those models of DE is just the number of
parameters. In our discussions, we have adopted  four widely used
parameterizations for the equation of state of DE in the literature.
For different numbers of bins, there are also common results.
Therefore our results are of some universality in some sense that
$\delta(z)$ prefers to cross the $\delta=0$ line and have an oscillation behavior.
In particular, this results indicate that if there does not exist any
interaction between DM and DE, the model of DE should be paid
special attention with an oscillating equation of state, because the
oscillating behavior of the interacting form is mathematically
equivalent to the case without interaction, but with an oscillating
equation of state of DE.

{\bf Acknowledgments:} This work was supported in part by the
National Natural Science Foundation of China under Grant Nos.
10535060, 10821504 and 10975168, and by National Basic Research
Program of China under Grant No. 2010CB833004.

\newpage
\begin{sidewaystable}[t]
\caption{Constraints of the interaction between DE and DM.
 For each parametrization of DE, the first line refers to positions of Maximum Likelihoods and errors of parameters at 68\% and 95\% c.l. ,
 the second line gives the best-fitted values of the parameters, in which we have set $\delta(z>1.8)=\delta_4=0$.
 The third line is for the best-fitted models with $\delta_4$ unfixed.
The value in \{\} means this parameter has been fixed.}
  \small
 \begin{tabular}{c|ccccccccc}
 \hline
 \hline
EoS of DE & & h & $\Omega_m^0$ & $w_0$ & $w_1$ & $\delta_1$ & $\delta_2$ & $\delta_3$ & $\delta_4$\\
 \hline
 I.$w=-1$ ($\delta_4=0$) & ML & $0.73^{+0.02+0.05}_{-0.02-0.06}$&$0.28^{+0.04+0.08}_{-0.06-0.12}$&\{-1\}&\{0\}&$0.13^{+0.22+0.42}_{-0.19-0.41}$&$-0.79^{+0.43+0.92}_{-0.40-0.61}$&$0.56^{+0.45+0.86}_{-0.59-0.90}$&\{0\}\\
 &best-fitted&0.734&0.244&\{-1\}&\{0\}&0.127&-0.804&0.770&\{0\}\\
\hline
$\delta_4$ unfixed&best-fitted&0.735&0.246&\{-1\}&\{0\}&0.133&-0.810&0.778&-0.097\\
 \hline
 II. $w=w_0$ ($\delta_4=0$) &ML&$0.73^{+0.03+0.06}_{-0.02-0.05}$&$0.28^{+0.04+0.08}_{-0.05-0.13}$&$-0.86^{+0.11+0.18}_{-0.28-0.72}$&\{0\}&$0.23^{+0.28+0.47}_{-0.41-0.74}$&$-0.85^{+0.75+1.33}_{-0.47-0.77}$&$0.50^{+0.45+1.00}_{-0.60-0.88}$&\{0\}\\
 &best-fitted&0.735&0.204&-0.792&\{0\}&0.484&-1.545&1.107&\{0\}\\
\hline
 $\delta_4$ unfixed&best-fitted&0.735&0.218&-0.797&\{0\}&0.518&-1.558&1.128&-0.603\\
 \hline
 III.\ \ \ \ \  ($\delta_4=0$)\ \ \ \ \  &ML&$0.73^{+0.03+0.05}_{-0.02-0.05}$&$0.29^{+0.05+0.08}_{-0.06-0.11}$&$-0.97^{+0.22+0.38}_{-0.29-0.75}$&$1.22^{+0.37+0.75}_{-1.81-2.75}$&$0.29^{+0.27+0.44}_{-0.45-0.79}$&$-0.59^{+0.67+1.26}_{-0.64-1.04}$&$0.51^{+0.59+1.15}_{-0.60-1.03}$&\{0\}\\
$ w=w_0+w_1z/(1+z)$&best-fitted&0.733&0.242&-0.632&-1.507&0.574&-1.802&1.134&\{0\}\\
\hline
 $\delta_4$ unfixed&best-fitted&0.734&0.232&-0.627&-1.533&0.551&-1.801&1.131&-0.388\\
 \hline
 IV.\ \ \ \ \   ($\delta_4=0$)\ \ \ \ \  &ML&$0.73^{+0.02+0.05}_{-0.02-0.06}$&$0.30^{+0.04+0.08}_{-0.05-0.14}$&$-0.98^{+0.65+1.67}_{-0.42-0.94}$&$2.9^{+2.0+4.1}_{-12.2-38.3}$&$-0.29^{+0.57+0.89}_{-0.34-0.55}$&$0.02^{+0.52+0.86}_{-0.82-1.53}$&$0.58^{+0.38+0.81}_{-0.53-1.39}$&\{0\}\\
 $w=w_0+w_1z/(1+z)^2 $ &best-fitted&0.733&0.221&-0.594&-2.115&0.524&-1.807&1.288&\{0\}\\
\hline
$\delta_4$ unfixed&best-fitted&0.733&0.228&-0.598&-2.098&0.540&-1.805&1.287&-0.277\\
 \hline
 \hline
\end{tabular}\label{ta}
\end{sidewaystable}
\end{document}